\documentclass[sigconf]{acmart}

\usepackage{booktabs}
\usepackage{graphicx}
\usepackage{url}
\usepackage{xcolor}
\usepackage{multirow}
\usepackage{enumitem}
\usepackage{balance}

\copyrightyear{2026}
\acmYear{2026}
\setcopyright{cc}
\setcctype{by}
\acmConference[CIKM '26]{Proceedings of the 35th ACM International Conference on Information and Knowledge Management}{November 07--11, 2026}{Rome, Italy}
\acmBooktitle{Proceedings of the 35th ACM International Conference on Information and Knowledge Management (CIKM '26), November 07--11, 2026, Rome, Italy}
\acmDOI{10.1145/3799682.3840771}
\acmISBN{979-8-4007-2539-5/2026/11}
\begin{document}

\title{Static Pruning Across Sparse Retrieval Regimes:\\What Transfers, What Breaks, and What Still Helps}

\author{Zirui Song}
\orcid{0009-0004-7984-1626}
\affiliation{%
  \institution{Amazon Web Services}
  \city{Shanghai}
  \country{China}}
\email{zrsong@amazon.com}

\author{Yuye Zhu}
\orcid{0009-0007-0746-1837}
\affiliation{%
  \institution{Amazon Web Services}
  \city{Shanghai}
  \country{China}}
\email{yuyezhu@amazon.com}

\author{Yang Yang}
\orcid{0009-0005-2480-3339}
\affiliation{%
  \institution{Amazon Web Services}
  \city{Shanghai}
  \country{China}}
\email{yych@amazon.com}

\begin{abstract}
Static pruning is widely used to accelerate sparse neural retrieval, yet existing studies each validate their conclusions within a single custom pipeline, leaving it unclear which findings transfer to modern engines with different index organizations and dynamic pruning mechanisms.
We present the first cross-engine pruning portability study, evaluating static pruning strategies across three engines---a controlled C++ pipeline (exhaustive inverted index), BMP (block-max pruning), and SEISMIC (clustered inverted indexes)---on two benchmarks (MS~MARCO, Natural Questions) with two encoders spanning opposite query-density regimes (SPLADE: 44 avg.\ query terms; V3-GTE: 7 avg.\ query terms), totaling 1{,}140 experimental configurations, with an additional deep-judgment validation on TREC~DL~2019/2020.
We find that \emph{index-side pruning (document and posting-list) is portable}: it consistently reduces latency (1.2--6.6$\times$) and index size (18--82\%) across all engines because sparse retrieval is memory-bound---a conclusion we support with cache-miss, TLB, and IPC profiling.
In contrast, \emph{query pruning is already internalized} by modern engines: it yields 4--11$\times$ speedup on the exhaustive pipeline but is subsumed by BMP's $\beta$ and SEISMIC's \texttt{query\_cut}.
Static pruning \emph{complements} dynamic pruning: on BMP, combining document and query reduction yields 2.5$\times$ speedup with NDCG@10 within 0.003 of the exact baseline.
Finally, NDCG@10 saturates while Recall@10 is still in the ${\sim}$85--95\% range across all three engines, providing a portable stopping criterion: practitioners can push pruning to this knee without visible ranking degradation. Together, these findings answer \emph{what transfers} (index-side pruning), \emph{what breaks} (query pruning), and \emph{what still helps} (static atop dynamic pruning).
Code is available at \url{https://github.com/zirui-song-18/cross_engine_static_pruning}.
\end{abstract}


\maketitle

\section{Introduction}
\label{sec:intro}

Sparse neural retrieval integrates neural text encoders with the inverted index, enabling term-based lookup while capturing semantic matching through learned sparse representations~\cite{formal2024towards}.
Models such as SPLADE~\cite{formal2021splade}, DeepImpact~\cite{mallia2021learning}, and uniCOIL~\cite{lin2021few} assign high-dimensional sparse weight vectors to queries and documents, where non-zero entries and their magnitudes can be interpreted as learned impacts.
Compared with classic bag-of-words retrieval, learned sparse models improve semantic matching but often increase practical costs: they activate substantially more query terms, each posting carries a real-valued learned weight, and the resulting impact distributions deviate from the sharply Zipfian patterns that traditional top-$k$ optimizations exploit~\cite{bruch2024efficient}.
Na\"ive deployments of these models on standard inverted-index engines can incur substantially higher latency than BM25 baselines~\cite{hofstatter2019let,macavaney2019cedr}.
Static pruning---removing low-weight query terms online, or low-impact document/posting entries from the index offline---is widely adopted to close this efficiency gap~\cite{lassance2023static}, shrinking the index and the per-query working set.
But choosing \emph{what} to prune, \emph{how aggressively}, and whether the chosen strategy remains effective across different retrieval engines are all first-order deployment decisions that the literature has not jointly addressed.

Lassance et al.~\cite{lassance2023static} provided the most comprehensive taxonomy, showing that learned sparse indexes tolerate aggressive pruning because their impact distributions are less sharply Zipfian.
However, all existing studies validate on custom pipelines with exhaustive scoring---no work has tested whether these conclusions transfer to modern engines with built-in dynamic pruning.
This gap is consequential: BMP~\cite{mallia2024faster} exposes a query-term fraction $\beta$ that itself acts as query pruning, and SEISMIC~\cite{bruch2024efficient} limits active terms via \texttt{query\_cut}.
Both engines already internalize query-term selection, yet no prior work has tested whether external static pruning adds value atop these mechanisms, whether static and dynamic pruning are complementary or redundant, or whether NDCG@10's saturation before Recall@10 generalizes beyond custom pipelines.

We present the first \emph{cross-engine pruning portability study} for sparse neural retrieval.
We systematically evaluate query, document, and posting-list pruning under two score-aware criteria ($\alpha$-Mass and Max-Ratio) across \textbf{three engines} (a controlled C++ pipeline with window-switch accumulator, BMP, and SEISMIC), \textbf{two datasets} (MS~MARCO with 8.8M passages and Natural Questions with 2.7M passages), and \textbf{two encoders} spanning distinct query-density regimes (SPLADE with $\sim$44 query terms and V3-GTE with $\sim$6.9 query terms)---\textbf{1,140 experimental configurations} in total.
A unified memory-bound thesis connects all three research questions: sparse retrieval is fundamentally limited by memory traffic, and this single fact explains which pruning transfers across engines, why static pruning complements dynamic pruning, and why NDCG saturates before Recall.
Our three interconnected contributions are:

\noindent\textbf{(1) Portability analysis} (Section~\ref{sec:rq1}).
Index-side pruning (document and posting-list) is \emph{portable across all tested engines}---it reduces both index size and latency consistently, with document pruning as the safest default.
Query pruning is \emph{regime-dependent}: largely redundant with engines that already internalize query-term selection.

\noindent\textbf{(2) Static--dynamic complementarity} (Section~\ref{sec:rq2}).
Static and dynamic pruning target orthogonal bottlenecks and yield combined speedups exceeding individual gains---a conclusion supported by micro-architectural profiling confirming the memory-bound thesis.

\noindent\textbf{(3) Portable stopping criterion} (Section~\ref{sec:metrics}).
NDCG@10 saturates at ${\sim}$85--95\% Recall@10 (engine-dependent) across all three tested engines, providing a portable empirical stopping criterion that we validate under both shallow and deep (TREC~DL) judgments.

\section{Background and Related Work}
\label{sec:related}

\paragraph{Sparse Neural Retrieval Models.}
Sparse neural retrievers encode queries and documents as high-dimensional sparse vectors whose non-zero entries carry learned impact weights~\cite{formal2024towards}.
Compared with classic bag-of-words scoring, learned sparse representations often activate substantially more terms due to expansion, and the resulting posting weights deviate from the traditional tf-idf distribution~\cite{bruch2024efficient}.
\emph{Bi-encoder} models such as SPLADE~\cite{formal2022distillation} apply learned expansion to both queries and documents, producing moderately dense representations (SPLADE: avg.\ 44 query terms, 120 doc terms on MS~MARCO).
\emph{Inference-free} encoders such as V3-GTE~\cite{geng2024towards} use tokenization-only query construction (avg.\ 6.9 terms) but expand documents more aggressively (avg.\ 180 terms).
These contrasting sparsity profiles create fundamentally different retrieval workloads: the dense-query regime generates more lists to traverse per query, while the sparse-query regime concentrates score mass on fewer terms, making each query term more critical.
Our study spans both regimes to disentangle encoder-specific artifacts from engine-portable patterns.

\paragraph{Static Pruning for Learned Sparse Retrieval.}
Static pruning has a long history in IR, spanning term-centric approaches that remove low-utility vocabulary terms~\cite{carmel2001static,blanco2007static}, document-centric strategies that prune per-document entries~\cite{buttcher2006document,de2005improving}, and global methods that remove postings based on corpus-level statistics~\cite{ntoulas2007pruning,altingovde2012static}.
Probabilistic and information-theoretic accounts have further refined the understanding of when pruning preserves retrieval quality~\cite{blanco2010probabilistic,chen2013information}.
For learned sparse retrievers specifically, Lassance et al.~\cite{lassance2023static} revisited these families and showed that aggressive pruning is feasible with two-stage pipelines, and that learned sparse indexes tolerate more aggressive pruning than traditional indexes because their impact distributions are less sharply Zipfian.
However, their study---like all prior pruning work for learned sparse models---validates on a custom pipeline with exhaustive scoring, not on modern engines with built-in dynamic pruning.
Our work extends this framework to BMP and SEISMIC, testing whether the conclusions are portable or pipeline-specific.

\paragraph{Dynamic Pruning and Modern Engines.}
Dynamic pruning avoids scoring documents that cannot enter the top-$k$: WAND~\cite{broder2003efficient} uses term-wise upper bounds, Block-Max WAND~\cite{ding2011faster} refines with block-level bounds, and impact-ordered designs enable early termination~\cite{anh2006pruned}.
These techniques underpin production systems (Lucene~\cite{grand2020maxscore}, PISA~\cite{mallia2019pisa}) and specialized learned-sparse engines~\cite{mallia2022faster,qiao2023optimizing,mackenzie2022accelerating}.
\textbf{BMP} \cite{mallia2024faster} extends Block-Max WAND with parameters $\alpha$ (approximation quality) and $\beta$ (query-term fraction---prunes terms with weight $< \beta \cdot w_{\max}$); reducing $\beta$ from 1.0 to 0.5 cuts latency by ${\sim}1.8\times$ on dense-query workloads~\cite{mallia2024faster}.
\textbf{SEISMIC}~\cite{bruch2024efficient} clusters posting lists with quantized summaries, exposing \texttt{query\_cut} (qc) and \texttt{heap\_factor} (hf); at qc\,=\,5 it achieves sub-millisecond retrieval.
DSP~\cite{carlson2025dynamic} and SINDI~\cite{li2025sindi} represent further designs; we select BMP and SEISMIC as exemplars of two dominant paradigms (block-max and clustered).
No prior work has systematically tested whether \emph{external} static pruning adds value atop these engines' built-in mechanisms.

\paragraph{Memory Locality and System-Level Acceleration.}
Score accumulation is a sparse vector operation with low arithmetic intensity and irregular access patterns~\cite{zhang2008performance}.
Modern CPU deployments are often limited by memory stalls---irregular accesses to postings and accumulator arrays---rather than raw arithmetic throughput, making locality-aware traversal central to speedups~\cite{mallia2019pisa,zhang2008performance}.
This memory-bound character is critical to understanding our cross-engine results: pruning that reduces memory traffic (e.g., document pruning that shortens posting lists) should transfer across engines regardless of their internal architecture, while pruning that reduces computation (e.g., query pruning that reduces the number of lists traversed) may not help when the engine is already memory-stalled.

\paragraph{Positioning.}
Prior work studies pruning \emph{within} a single engine~\cite{lassance2023static,mallia2024faster,bruch2024efficient}.
We study pruning \emph{across} engines, yielding a portability matrix that replaces engine-specific heuristics with cross-validated guidance.

\section{Pruning Strategies}
\label{sec:pruning}

We formalize two score-aware pruning criteria and three pruning families.
We omit Fixed-Top selection, whose integer cutoffs prune objects of different lengths unevenly and cannot produce the continuous parameter sweeps that $\alpha$-Mass and Max-Ratio provide.

\subsection{Pruning Criteria}
\label{sec:criteria}

Let $\mathcal{T}(x)$ denote the set of non-zero terms in a sparse object $x$ with weights $\{w_t\}_{t \in \mathcal{T}(x)}$
(where $x = q$ for a query or $x = d$ for a document). For posting-list pruning, the same criteria apply over the posting weights $w_{t,d}$ within each list $L_t$, selecting a subset of documents to retain.
We define two criteria that select a pruned support $\mathcal{T}^{\mathrm{pruned}}(x) \subseteq \mathcal{T}(x)$:

\paragraph{$\alpha$-Mass (AM)}
Sort weights in descending order and keep the smallest prefix reaching $\alpha$ of the $\ell_1$ mass:
$\mathcal{T}^{\mathrm{pruned}}(x) = \{t_1, \ldots, t_m\}$ such that
$\sum_{i=1}^{m} w_{t_i} \ge \alpha \sum_{t \in \mathcal{T}(x)} w_t$,
with $\alpha \in (0,1]$.
This criterion yields adaptive support size---high-entropy objects retain more terms.

\paragraph{Max-Ratio (MR)}
Keep terms whose weight is at least a fraction $\tau$ of the maximum:
$\mathcal{T}^{\mathrm{pruned}}(x) = \{t \in \mathcal{T}(x) \mid w_t \ge \tau \cdot w_{\max}\}$,
where $w_{\max} = \max_{t \in \mathcal{T}(x)} w_t$ and $\tau \in [0,1)$.
(We use $\tau$ to distinguish from BMP's $\beta$ parameter.)
This criterion is scale-invariant and produces continuous trade-off curves.

\subsection{Pruning Families}
\label{sec:families}

\paragraph{Query pruning (online).}
\textbf{Pruned object:} query vector $q$ (support $\mathcal{T}(q)$).
\textbf{Mechanism:} apply a criterion to obtain $\mathcal{T}^{\mathrm{pruned}}(q)$ and traverse only the corresponding posting lists.
\textbf{Cost path reduced:} number of lists traversed (and thus accumulator updates).
Query pruning is performed on-the-fly per query and does not modify the index.
On engines with built-in query selection (BMP's $\beta$, SEISMIC's \texttt{query\_cut}), external query pruning may be partially or fully redundant.

\paragraph{Document pruning (offline, per-document).}
\textbf{Pruned object:} each document vector $d$ independently prior to indexing~\cite{lassance2023static}.
\textbf{Mechanism:} apply a criterion to obtain $\mathcal{T}^{\mathrm{pruned}}(d)$ and build the inverted index from pruned documents.
\textbf{Cost path reduced:} per-list work and RAM footprint (fewer postings overall), shrinking the working set.
Document pruning is irreversible at query time.
Because it reduces memory traffic rather than computation, we hypothesize that it transfers across all engine architectures.

\paragraph{Posting-list pruning (offline, per-term).}
\textbf{Pruned object:} each posting list $L_t$ independently.
\textbf{Mechanism:} apply a criterion to keep only high-impact postings in $L_t$~\cite{bruch2024efficient}.
\textbf{Cost path reduced:} per-list work and RAM (shorter lists), yielding large savings when queries hit frequent terms.
Like document pruning, deletions are irreversible.
Posting-list pruning achieves the largest RAM savings but exhibits more rapid Recall degradation under aggressive settings than document pruning.

\section{Experimental Setup}
\label{sec:setup}

\paragraph{Datasets.}
We evaluate on two benchmarks:
\textbf{MS~MARCO} passage retrieval~\cite{nguyen2016ms} (${\sim}$8.8M passages, 6,980 dev queries), the standard sparse retrieval benchmark with shallow relevance judgments (1--2 per query); and
\textbf{Natural Questions (NQ)}~\cite{thakur2021beir} from BEIR (${\sim}$2.7M passages, 3,452 queries), providing cross-dataset validation under different corpus size and judgment characteristics.

\paragraph{Encoders.}
We evaluate two encoders with opposite query-density regimes (Table~\ref{tab:dataset_stats}): \textbf{SPLADE-CoCondenser-EnsembleDistil}~\cite{formal2022distillation}, a bi-encoder that expands both queries and documents (\emph{dense-query} regime), and \textbf{opensearch-neural-sparse-encoding-doc-v3-gte} (V3-GTE)~\cite{geng2024towards}, an inference-free encoder with tokenized-only queries but aggressive document expansion (\emph{sparse-query} regime). This lets us disentangle query density from encoder architecture.

\begin{table}[h]
\centering
\caption{Dataset and encoder statistics. Avg.\ non-zero counts (nnz) per query and document, corpus size, baseline NDCG@10 on the controlled pipeline, and baseline NDCG@10 on the deep-judgment TREC~DL~2019/2020 topics (mean over both years; MS~MARCO corpus).}
\label{tab:dataset_stats}
\small
\setlength{\tabcolsep}{1.5pt}
\begin{tabular}{l c c c c c}
\toprule
Configuration & Query nnz & Doc nnz & Docs & NDCG@10 & DL'19/20 \\
\midrule
MS~MARCO + SPLADE    & 43.95  & 119.96 & 8.8M & 0.449 & 0.726 \\
MS~MARCO + V3-GTE    &  6.89  & 180.41 & 8.8M & 0.428 & 0.720 \\
NQ + SPLADE           & 46.97  & 147.52 & 2.7M & 0.539 & --- \\
NQ + V3-GTE           &  7.20  & 185.00 & 2.7M & 0.582 & --- \\
\bottomrule
\end{tabular}
\end{table}

\paragraph{Engine 1: Controlled C++ Pipeline.}
We build a single-threaded, core-pinned inverted-index engine in C++ implementing a \emph{window-switch accumulator} ($\Psi$).
Documents are partitioned into windows of size $W$; postings are stored per (term, window) pair with local IDs, keeping the accumulator cache-resident at $O(W)$ rather than $O(N)$.
Queries score exhaustively against the (possibly pruned) index, then optionally re-rank the top-$k'=50$ candidates using the full (unpruned) document vectors.
Since BMP and SEISMIC operate single-stage, Appendix~\ref{app:rerank_ablation} reports a no-rerank ablation confirming that the qualitative portability findings are not driven by the re-ranker.
Section~\ref{sec:rq2_mechanism} and Appendix~\ref{app:accumulators} introduce two additional accumulator variants ($\Phi$, $\Xi$) designed to isolate the memory-bound bottleneck via controlled profiling.

\paragraph{Engine 2: BMP}
Block-max dynamic pruning engine~\cite{mallia2024faster}.
Indexes are built from CIFF format with 8-bit quantized impacts.
Parameters: $\alpha$ (approximation quality) and $\beta$ (query-term fraction---retains terms with weight $\ge \beta \cdot w_{\max}$).
BMP's $\beta$ parameter is itself a query-pruning mechanism; we test whether external static pruning adds value beyond it.

\paragraph{Engine 3: SEISMIC}
Clustered inverted-index engine~\cite{bruch2024efficient}.
Parameters: \texttt{query\_cut} (qc, limits active query terms) and \texttt{heap\_factor} (hf, controls dynamic skipping aggressiveness).
SEISMIC achieves sub-millisecond retrieval through clustering and quantized summaries; its \texttt{query\_cut} already incorporates query-term selection, analogous to BMP's $\beta$.

\paragraph{Metrics.}
We report four metrics serving complementary purposes.
\textbf{Recall@$k$}: oracle fidelity ($|A_k \cap G_k| / k$, where $G_k$ is the model's exact top-$k$).
\textbf{NDCG@$k$}: qrels-based ranking quality with top-heavy logarithmic discounting~\cite{jarvelin2002cumulated}.
\textbf{Success@$k$}: qrels-based coverage ($\mathbb{I}[A_k \cap R(q) \ne \emptyset]$, where $R(q)$ is the set of judged-relevant documents).
\textbf{Latency}: mean (BMP/SEISMIC); mean, p95, and p99 (controlled pipeline).
Index size is reported as in-memory inverted-index footprint (GB).

\paragraph{Protocol.}
All experiments run single-threaded with CPU core pinning on an AMD EPYC 9R14 processor (3.7\,GHz, 1.5\,TB RAM).
Each configuration uses 5 warm-up runs followed by timed runs; we report mean latency.
Default $k = 10$ for all metrics unless stated.


\section{Portability of Pruning Strategies}
\label{sec:rq1}

\begin{figure*}[h]
\centering
\includegraphics[width=0.95\linewidth]{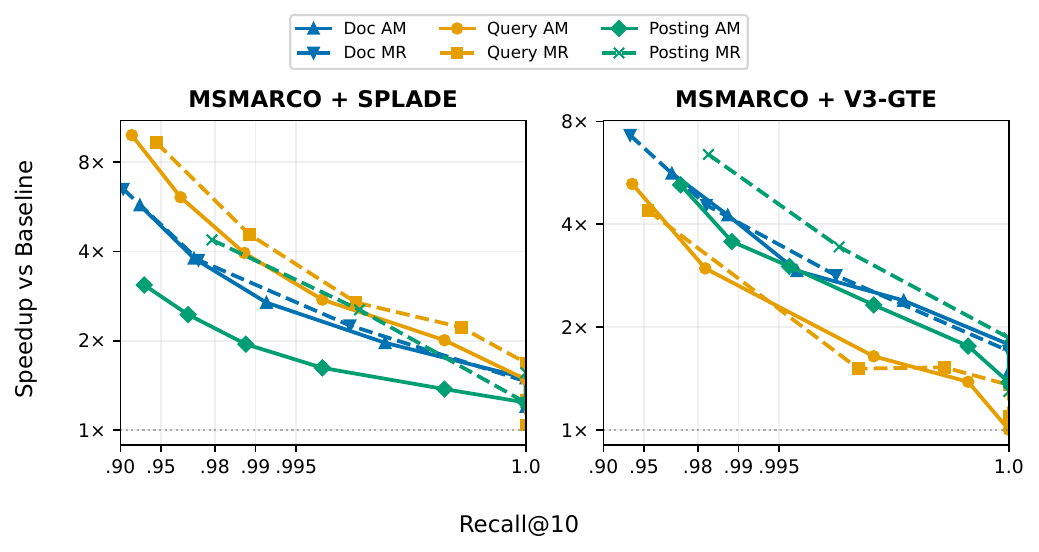}
\caption{Custom C++ Pareto frontiers on MS~MARCO: speedup vs.\ $\text{Recall@10}$ for all six pruning families (\{Doc, Query, Posting\} $\times$ \{Alpha-Mass, Max-Ratio\}).
\textbf{Left}: SPLADE. \textbf{Right}: V3-GTE. Document pruning provides the most stable high-recall frontier across both encoders.}
\label{fig:pareto_cross}
\Description{Six Pareto curves comparing pruning families on two encoder regimes. Document pruning provides the most stable frontier in both.}
\end{figure*}

\textbf{RQ1:} \emph{Which static pruning conclusions are portable across engine designs and query-density profiles?}
\textbf{Answer:} Index-side pruning (document and posting-list) transfers across all tested engines because it reduces memory traffic---the binding bottleneck---whereas query pruning is regime-dependent, subsumed by engines that already internalize query-term selection.
Figure~\ref{fig:pareto_cross} previews this on the controlled C++ pipeline: document pruning provides the most stable high-recall frontier across both encoders, a finding confirmed by the engine-specific results (Sections~\ref{sec:rq1_bmp}--\ref{sec:rq1_seismic}).

\subsection{Controlled Pipeline Results}
\label{sec:rq1_controlled}

\paragraph{Baselines.}
Table~\ref{tab:controlled_ops} reports baseline latencies, ranging from 11.3\,ms (NQ+V3-GTE) to 70.1\,ms (MS+SPLADE), all at Recall@10\,=\,1.0 (exhaustive scoring).

\paragraph{Query pruning is highly effective on exhaustive pipelines.}
Query $\alpha$-Mass pruning yields 4--11$\times$ speedups on the controlled pipeline (Table~\ref{tab:controlled_ops}), but with a regime-dependent cost: SPLADE retains near-perfect NDCG (0.448 vs.\ 0.449) at $\alpha$\,=\,0.50, while V3-GTE suffers steep Recall loss (0.775) because its ${\sim}$7 query terms each carry critical score mass.
The key question is whether these gains survive on engines with \emph{built-in} query-term selection.

\paragraph{Document pruning is consistently effective.}
Document pruning achieves nearly identical speedup ratios across SPLADE and V3-GTE (Table~\ref{tab:controlled_ops}), confirming it targets \emph{memory traffic} rather than per-query computation---making it encoder-agnostic.

\paragraph{Posting-list pruning: RAM-efficient but faster degradation.}
Posting pruning achieves comparable speedups to document pruning (Table~\ref{tab:controlled_ops}) with even larger index reductions (73--81\%).
It degrades Recall faster, however: document pruning keeps each document's own top-weighted terms, whereas posting-list pruning thresholds each term globally, so a document whose weight for a query's key term falls just below the cutoff is dropped entirely---even when relevant. It thus suits RAM-constrained deployments, with the threshold tuned under a validation constraint.

\begin{table}[h]
\caption{Controlled pipeline ($\Psi$): representative operating points with
two-stage re-rank ($k'$\,=\,50). Speedup relative to unpruned baseline. Idx is the in-memory inverted-index size.}
\label{tab:controlled_ops}
\centering
\small
\resizebox{\linewidth}{!}{%
\begin{tabular}{@{}l l r r r r r r@{}}
\toprule
\textbf{Dataset+Enc.} & \textbf{Pruning} & \textbf{Lat.\,(ms)} & \textbf{Spd.} & \textbf{Idx\,(GB)} & \textbf{R@10} & \textbf{NDCG} & \textbf{MRR} \\
\midrule
\multirow{5}{*}{MS+SPL}
 & Baseline          & 70.1  & 1.0$\times$ & 8.15 & 1.000 & 0.449 & 0.383 \\
 & Q AM\,0.50         & 11.5  & 6.1$\times$ & 8.15 & 0.964 & 0.448 & 0.383 \\
 & Doc AM\,0.50       & 11.7  & 6.0$\times$ & 1.53 & 0.928 & 0.443 & 0.379 \\
 & Doc MR\,0.30       & 18.1  & 3.9$\times$ & 2.16 & 0.974 & 0.447 & 0.381 \\
 & Post MR\,0.30      & 15.6  & 4.5$\times$ & 2.00 & 0.979 & 0.447 & 0.382 \\
\midrule
\multirow{5}{*}{MS+GTE}
 & Baseline          & 29.5  & 1.0$\times$ & 12.23 & 1.000 & 0.428 & 0.362 \\
 & Q AM\,0.50         &  2.6  & 11.3$\times$ & 12.23 & 0.775 & 0.389 & 0.332 \\
 & Doc AM\,0.50       &  5.2  & 5.6$\times$ & 2.32 & 0.969 & 0.426 & 0.360 \\
 & Doc MR\,0.30       &  6.2  & 4.8$\times$ & 2.90 & 0.983 & 0.427 & 0.361 \\
 & Post MR\,0.30      &  4.4  & 6.7$\times$ & 2.29 & 0.983 & 0.427 & 0.361 \\
\midrule
\multirow{5}{*}{NQ+SPL}
 & Baseline          & 29.0  & 1.0$\times$ & 3.04 & 1.000 & 0.539 & 0.488 \\
 & Q AM\,0.50         &  6.8  & 4.3$\times$ & 3.04 & 0.972 & 0.537 & 0.487 \\
 & Doc AM\,0.50       &  5.7  & 5.1$\times$ & 0.58 & 0.894 & 0.530 & 0.482 \\
 & Doc MR\,0.30       & 10.2  & 2.8$\times$ & 0.81 & 0.962 & 0.537 & 0.487 \\
 & Post MR\,0.30      & 10.5  & 2.8$\times$ & 0.84 & 0.978 & 0.538 & 0.487 \\
\midrule
\multirow{5}{*}{NQ+GTE}
 & Baseline          & 11.3  & 1.0$\times$ & 4.26 & 1.000 & 0.582 & 0.535 \\
 & Q AM\,0.50         &  1.5  & 7.4$\times$ & 4.26 & 0.800 & 0.550 & 0.511 \\
 & Doc AM\,0.50       &  3.0  & 3.7$\times$ & 0.84 & 0.969 & 0.579 & 0.533 \\
 & Doc MR\,0.30       &  2.5  & 4.5$\times$ & 1.09 & 0.983 & 0.580 & 0.534 \\
 & Post MR\,0.30      &  2.5  & 4.5$\times$ & 0.91 & 0.980 & 0.580 & 0.534 \\
\bottomrule
\end{tabular}
}
\end{table}

\subsection{BMP Validation}
\label{sec:rq1_bmp}

\paragraph{BMP baselines.}
BMP baselines range from 1,087\,$\mu$s (NQ+V3-GTE) to 8,915\,$\mu$s (MS+SPLADE) at exact retrieval (Table~\ref{tab:portability_matrix}).

\paragraph{BMP's $\beta$ is the dominant query lever.}
BMP's internal $\beta$ already \emph{internalizes} the same query-term selection that yields 4--11$\times$ on the pipeline.
External static query pruning (MR\,0.10) provides 1.9$\times$ on MS+SPLADE (Table~\ref{tab:portability_matrix})---partially overlapping with $\beta$'s effect rather than adding a new independent optimization dimension.

\paragraph{Document pruning transfers to BMP}
Document pruning reduces the BMP index ($-$34--36\%) and latency (1.2--1.4$\times$) with NDCG@10 within 0.007 of the unpruned baseline (Table~\ref{tab:portability_matrix}).
The mechanism is clear: shorter posting lists tighten BMP's block-max upper bounds, improving both cache locality and dynamic skipping efficiency.
The pattern holds across both datasets and encoders.

\paragraph{Posting pruning also transfers.}
Posting pruning provides index reduction comparable to document pruning ($-$34--37\%) with similar latency gains on BMP (Table~\ref{tab:portability_matrix}).
Consistent with controlled-pipeline findings, it degrades Recall faster than document pruning under aggressive settings.

\paragraph{V3-GTE on BMP: document pruning is the safer choice.}
On V3-GTE, $\beta$\,=\,0.5 causes steep Recall degradation (0.706--0.778) because each of the ${\sim}$7 query terms carries critical score mass.
In contrast, Doc~AM\,0.90 retains Recall@10\,$\ge$\,0.966 with 36\% index reduction (Table~\ref{tab:bmp_doc_detail}).
For sparse-query encoders, \emph{document pruning is the safer, more predictable choice} on BMP.

\subsection{SEISMIC Validation}
\label{sec:rq1_seismic}

\begin{figure}[h]
\centering
\includegraphics[width=0.95\linewidth]{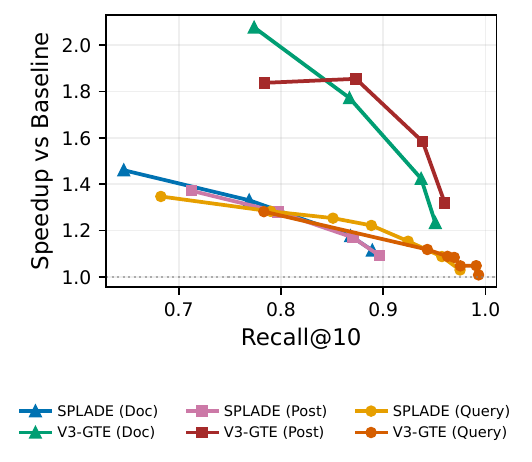}
\caption{SEISMIC Pareto frontiers (qc\,=\,5): speedup vs.\ Recall@10 under document and query pruning for both encoders. Document pruning provides consistent gains; query pruning overlaps with SEISMIC's internal \texttt{query\_cut}.}
\label{fig:seismic_pareto}
\Description{SEISMIC Pareto curves showing document pruning provides consistent gains while query pruning overlaps with internal query cut.}
\end{figure}

\paragraph{SEISMIC baselines: sub-millisecond retrieval.}
SEISMIC at qc\,=\,5 already achieves sub-millisecond retrieval (186--360\,$\mu$s across configurations), orders of magnitude faster than the controlled pipeline (Figure~\ref{fig:seismic_pareto}).

\paragraph{Query pruning provides marginal benefit.}
Static query MR\,0.10 at qc\,=\,5 on MS~MARCO reduces latency from 186 to
161\,$\mu$s but Recall@10 drops to 0.924.
At qc\,=\,20, MR\,0.10 provides 1.31$\times$ (413 $\to$ 315\,$\mu$s).
SEISMIC's \texttt{query\_cut} already limits active query terms, so external
pruning partially overlaps---analogous to BMP's $\beta$.
On V3-GTE, the same pattern holds: query MR\,0.10 at qc\,=\,5 yields
244\,$\mu$s vs.\ 255\,$\mu$s baseline, a marginal 1.05$\times$ gain.

\paragraph{Document pruning reduces index size with modest latency gain.}
Doc~AM\,0.90 yields 15--34\% latency reduction with 16--29\% index savings across all configurations (Table~\ref{tab:seismic_doc_detail}).
Latency gains are smaller than BMP's because SEISMIC is already near its memory-bound floor; the primary benefit is \textbf{index reduction} for RAM-constrained deployment.

\paragraph{Posting pruning also transfers to SEISMIC}
Posting MR\,0.10 provides comparable index reduction and speedup to document pruning on SEISMIC, with similar Recall at matched operating points (Table~\ref{tab:portability_matrix}).
Both index-side families yield consistent latency and RAM benefits across all four configurations.

\subsection{Cross-Engine Portability Matrix}
\label{sec:rq1_synthesis}

Table~\ref{tab:portability_matrix} synthesizes findings across all three
engines into a \emph{portability matrix}---the core contribution of RQ1.

\begin{table*}[h]
\caption{Portability Matrix: three pruning methods $\times$ four configurations, each evaluated on three engines.
Speedup is relative to each engine's unpruned baseline.
$\Delta$NDCG and $\Delta$MRR show absolute change vs.\ the exact baseline (negative = degradation); $\Delta$MRR tracks $\Delta$NDCG throughout.
C++ uses two-stage re-ranking ($k'$\,=\,50); C++$^\dagger$ shows single-stage (no re-ranking) for direct comparison with BMP/SEISMIC.
$^*$BMP external query MR at $\beta$\,=\,1.0.}
\label{tab:portability_matrix}
\centering
\small
\setlength{\tabcolsep}{1.8pt}
\resizebox{\linewidth}{!}{%
\begin{tabular}{@{}l @{\hspace{6pt}} l @{\hspace{8pt}} cccc @{\hspace{8pt}} cccc @{\hspace{8pt}} cccc @{\hspace{8pt}} cccc@{}}
\toprule
& & \multicolumn{4}{c}{\textbf{Speedup}} & \multicolumn{4}{c}{\textbf{Recall@10}} & \multicolumn{4}{c}{\textbf{$\Delta$NDCG@10}} & \multicolumn{4}{c}{\textbf{$\Delta$MRR@10}} \\
\cmidrule(lr){3-6} \cmidrule(lr){7-10} \cmidrule(lr){11-14} \cmidrule(lr){15-18}
\textbf{Config} & \textbf{Pruning} & \textbf{C++} & \textbf{C++$^\dagger$} & \textbf{BMP} & \textbf{SEIS.} & \textbf{C++} & \textbf{C++$^\dagger$} & \textbf{BMP} & \textbf{SEIS.} & \textbf{C++} & \textbf{C++$^\dagger$} & \textbf{BMP} & \textbf{SEIS.} & \textbf{C++} & \textbf{C++$^\dagger$} & \textbf{BMP} & \textbf{SEIS.} \\
\midrule
\multirow{3}{*}{\rotatebox{0}{\scriptsize MS+SPL}}
 & Doc AM\,0.90  & 1.49$\times$ & 1.48$\times$ & 1.43$\times$ & 1.18$\times$ & 1.000 & .946 & .948 & .868 & .000 & $-$.003 & $-$.003 & $-$.005 & .000 & $-$.003 & $-$.002 & $-$.005 \\
 & Post MR\,0.10 & 1.56$\times$ & 1.54$\times$ & 1.47$\times$ & 1.17$\times$ & 1.000 & .947 & .948 & .871 & .000 & $-$.002 & $-$.001 & $-$.004 & .000 & $-$.002 & $-$.002 & $-$.004 \\
 & Query MR\,0.10& 1.68$\times$ & 1.65$\times$ & 1.91$\times$$^*$ & 1.17$\times$ & 1.000 & .945 & .935 & .924 & .000 & .000 & .000 & +.001 & .000 & $-$.001 & .000 & .000 \\
\midrule
\multirow{3}{*}{\rotatebox{0}{\scriptsize MS+GTE}}
 & Doc AM\,0.90  & 1.78$\times$ & 1.61$\times$ & 1.23$\times$ & 1.42$\times$ & 1.000 & .969 & .969 & .937 & .000 & $-$.001 & $-$.001 & $-$.002 & .000 & .000 & $-$.001 & $-$.001 \\
 & Post MR\,0.10 & 1.86$\times$ & 1.68$\times$ & 1.27$\times$ & 1.58$\times$ & 1.000 & .970 & .969 & .939 & .000 & $-$.002 & $-$.002 & $-$.002 & .000 & $-$.001 & $-$.002 & $-$.002 \\
 & Query MR\,0.10& 1.36$\times$ & 1.27$\times$ & 1.19$\times$$^*$ & 1.03$\times$ & 1.000 & .982 & .980 & .975 & .000 & +.001 & +.001 & +.001 & .000 & +.001 & +.001 & +.001 \\
\midrule
\multirow{3}{*}{\rotatebox{0}{\scriptsize NQ+SPL}}
 & Doc AM\,0.90  & 1.37$\times$ & 1.34$\times$ & 1.33$\times$ & 1.24$\times$ & 1.000 & .935 & .936 & .853 & .000 & $-$.003 & $-$.002 & $-$.002 & .000 & $-$.002 & $-$.001 & .000 \\
 & Post MR\,0.10 & 1.34$\times$ & 1.32$\times$ & 1.30$\times$ & 1.20$\times$ & 1.000 & .954 & .952 & .865 & .000 & $-$.002 & $-$.002 & $-$.001 & .000 & .000 & $-$.001 & +.001 \\
 & Query MR\,0.10& 1.42$\times$ & 1.39$\times$ & 1.60$\times$$^*$ & 1.20$\times$ & 1.000 & .956 & .947 & .908 & .000 & $-$.005 & $-$.006 & $-$.004 & .000 & $-$.004 & $-$.006 & $-$.004 \\
\midrule
\multirow{3}{*}{\rotatebox{0}{\scriptsize NQ+GTE}}
 & Doc AM\,0.90  & 1.43$\times$ & 1.43$\times$ & 1.19$\times$ & 1.51$\times$ & 1.000 & .965 & .966 & .950 & .000 & $-$.005 & $-$.001 & $-$.007 & .000 & $-$.005 & $-$.004 & $-$.006 \\
 & Post MR\,0.10 & 1.49$\times$ & 1.49$\times$ & 1.22$\times$ & 1.57$\times$ & 1.000 & .966 & .966 & .951 & .000 & $-$.005 & $-$.005 & $-$.006 & .000 & $-$.005 & $-$.005 & $-$.006 \\
 & Query MR\,0.10& 1.17$\times$ & 1.17$\times$ & 1.13$\times$$^*$ & 1.08$\times$ & 1.000 & .990 & .989 & .984 & .000 & $-$.001 & $-$.002 & $-$.001 & .000 & $-$.002 & $-$.003 & $-$.001 \\
\bottomrule
\end{tabular}
}
\end{table*}

\paragraph{Synthesis.}
Table~\ref{tab:portability_matrix} shows that index-side pruning is the transferable family: document and posting-list pruning consistently reduce memory traffic with small NDCG loss, while query pruning is largely internalized by BMP's $\beta$ and SEISMIC's \texttt{query\_cut}.
The C++$^\dagger$ column further confirms that the cross-engine gap is not merely a re-ranker artifact: without re-ranking, C++ aligns closely with BMP, whereas SEISMIC exhibits larger Recall loss under the same threshold---indicating higher effective pruning severity under clustered traversal (Figure~\ref{fig:rerank_ablation}).

\subsection{Per-Query Latency Stability}

\begin{figure}[h!]
    \centering
    \includegraphics[width=0.9\linewidth]{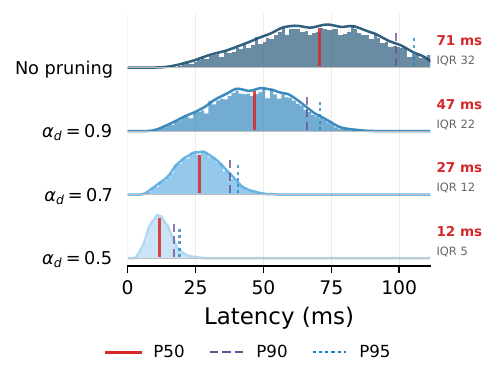}
    \caption{Per-query latency histogram strips on MS~MARCO+SPLADE (controlled pipeline) at four document-pruning levels. The P50/P90/P95 markers show progressive IQR compression: moderate pruning ($\alpha_d$\,=\,0.9) narrows the distribution while aggressive pruning ($\alpha_d$\,=\,0.5) concentrates it tightly with minimal tail degradation.}
    \label{fig:per_query_variance}
\Description{Histogram strips showing per-query latency distributions at four pruning levels with KDE outlines and percentile markers.}
\end{figure}

Using document pruning as a representative example, static pruning compresses the entire latency distribution rather than merely shifting the mean (Figure~\ref{fig:per_query_variance}): IQR drops substantially at $\alpha_d$\,=\,0.9, the P95/P50 ratio decreases monotonically, and no heavy tail emerges even at $\alpha_d$\,=\,0.5---aggressive pruning does not create a subpopulation of severely degraded queries.

\section{Static--Dynamic Complementarity}
\label{sec:rq2}

\textbf{RQ2:} \emph{Does static pruning complement or conflict with dynamic pruning?}
\textbf{Answer:} Static and dynamic pruning are complementary---they target orthogonal bottlenecks (memory footprint vs.\ wasted traversal), yielding combined gains that exceed either mechanism alone.

\subsection{BMP Factorial Design}
\label{sec:rq2_bmp}

We adopt a 2$\times$2 factorial design on BMP (MS~MARCO+SPLADE,
$\alpha$\,=\,1.0, $k$\,=\,10): \emph{full} vs.\ \emph{pruned} index
(Doc~AM\,0.90), crossed with \emph{full} vs.\ \emph{pruned} queries
(static MR\,0.10).  Table~\ref{tab:bmp_factorial} reports results.

\begin{table}[h]
\caption{BMP 2$\times$2 factorial on MS~MARCO+SPLADE ($\alpha$\,=\,1.0).
Speedup relative to the full-index, full-query baseline.
Index size in GB.}
\label{tab:bmp_factorial}
\centering
\small
\setlength{\tabcolsep}{3pt}
\resizebox{\linewidth}{!}{%
\begin{tabular}{@{}l l r r r r r r@{}}
\toprule
\textbf{Index} & \textbf{Query} & \textbf{Lat.\,($\mu$s)} & \textbf{Spd.} & \textbf{R@10} & \textbf{NDCG} & \textbf{MRR} & \textbf{Idx} \\
\midrule
Full  & Full       & 8,915 & 1.00$\times$ & 1.000 & 0.449 & 0.382 & 15.7 \\
Full  & Q MR\,0.10 & 4,680 & 1.91$\times$ & 0.935 & 0.449 & 0.382 & 15.7 \\
Pruned & Full      & 6,235 & 1.43$\times$ & 0.948 & 0.446 & 0.380 & 10.1 \\
Pruned & Q MR\,0.10& 3,537 & \textbf{2.52$\times$} & 0.917 & 0.446 & 0.379 & 10.1 \\
\bottomrule
\end{tabular}
}
\end{table}

\begin{figure}[h]
\centering
\includegraphics[width=0.85\linewidth]{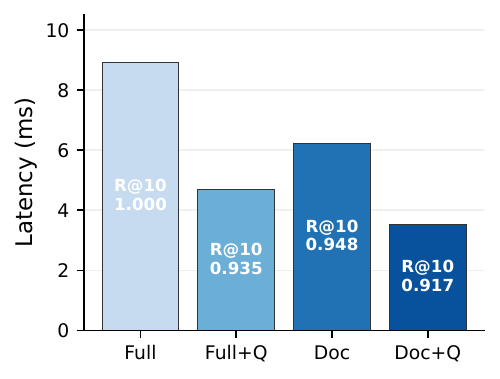}
\caption{BMP 2$\times$2 factorial on MS~MARCO+SPLADE: latency under four
conditions (full/pruned index $\times$ full/pruned queries).
Recall@10 shown inside each bar. Combined pruning achieves the largest speedup
(2.52$\times$) with minimal NDCG loss ($\Delta$NDCG\,=\,$-$0.003).}
\label{fig:bmp_factorial}
\Description{Bar chart showing four BMP conditions with latency bars and recall values. Combined pruning achieves the best speedup.}
\end{figure}

\paragraph{Combined pruning is stackable.}
The combined condition achieves \textbf{2.52$\times$} speedup (Table~\ref{tab:bmp_factorial}, Figure~\ref{fig:bmp_factorial}) with NDCG@10 within 0.003 of baseline.
The gain is submultiplicative ($\rho = 0.92$), indicating overlapping but not fully redundant cost components.
The two mechanisms target orthogonal dimensions: static pruning reduces memory footprint (shorter lists, better cache locality); dynamic pruning reduces wasted traversal (skips non-competitive blocks).
BMP's upper bounds become \emph{tighter} with shorter posting lists, yielding synergistic gains.
On NQ+SPLADE the pattern repeats (2.02$\times$); V3-GTE shows lower but still complementary gains due to less query-pruning headroom.

\subsection{SEISMIC Static--Internal Interaction}
\label{sec:rq2_seismic}

Table~\ref{tab:seismic_factorial} reports the SEISMIC factorial
(MS~MARCO+SPLADE, qc\,=\,5, hf\,=\,1.0).

\begin{table}[h]
\caption{SEISMIC on MS~MARCO+SPLADE (qc\,=\,5, hf\,=\,1.0).
Speedup relative to the unpruned baseline at qc\,=\,5. Index size in GB.}
\label{tab:seismic_factorial}
\centering
\small
\resizebox{\linewidth}{!}{%
\begin{tabular}{@{}l l r r r r r r@{}}
\toprule
\textbf{Index} & \textbf{Query} & \textbf{Lat.\,($\mu$s)} & \textbf{Spd.} & \textbf{R@10} & \textbf{NDCG} & \textbf{MRR} & \textbf{Idx} \\
\midrule
Full  & Full       & 185 & 1.00$\times$ & 0.977 & 0.443 & 0.379 & 8.1 \\
Full  & Q MR\,0.10 & 161 & 1.15$\times$ & 0.924 & 0.444 & 0.379 & 8.1 \\
Pruned & Full      & 158 & 1.17$\times$ & 0.868 & 0.438 & 0.374 & 6.0 \\
Pruned & Q MR\,0.10& 133 & \textbf{1.39$\times$} & 0.856 & 0.440 & 0.376 & 6.0 \\
\bottomrule
\end{tabular}
}
\end{table}

\paragraph{Complementary, but primarily through index reduction.}
Combined pruning yields 1.39$\times$ on MS+SPLADE (Table~\ref{tab:seismic_factorial}), smaller than BMP's 2.52$\times$ because SEISMIC's sub-millisecond baseline is near the memory-bound floor---the latency set by unavoidable index traversal once data movement is minimized. With little memory traffic left to eliminate, pruning's primary benefit becomes RAM reduction ($-$26\%) rather than speedup.

\subsection{Mechanistic Explanation: Memory-Bound Retrieval}
\label{sec:rq2_mechanism}

\begin{figure}[h]
\centering
\includegraphics[width=0.98\linewidth]{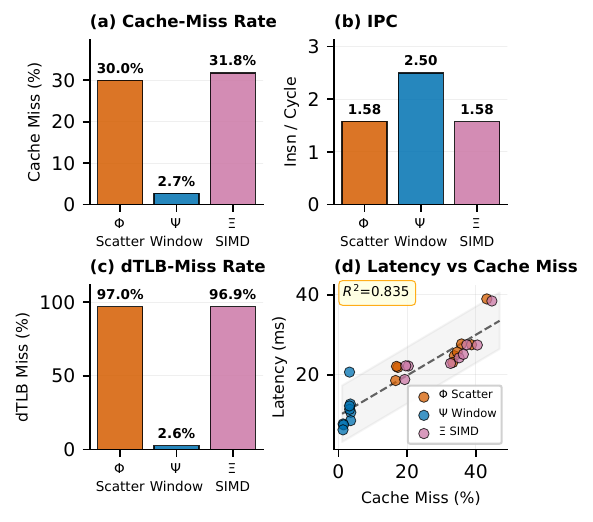}
\caption{Micro-architectural profiling of $\Phi$/$\Psi$/$\Xi$ with Linux
perf on multiple pruning configurations: (a) cache-miss rate,
(b) IPC, (c) dTLB-miss rate, (d) per-query latency vs.\ cache-miss rate
($R^2$\,=\,0.84).  Latency is tightly coupled to memory locality, not
arithmetic throughput.}
\label{fig:profiling}
\Description{Four-panel profiling figure: bar charts for cache-miss rate, IPC, and dTLB-miss rate, plus a scatter plot of latency versus cache-miss rate with regression line.}
\end{figure}

The factorial experiments above demonstrate that static and dynamic pruning are complementary, but do not explain \emph{why}.
If sparse retrieval were compute-bound, reducing arithmetic work (query pruning) and reducing wasted computation (dynamic skipping) would overlap.
If it is memory-bound, shrinking the working set (static document pruning) and avoiding unnecessary block reads (dynamic pruning) would target independent cost components.
We resolve this with controlled micro-architectural profiling, holding the index and queries fixed while varying only the accumulator architecture:
\begin{itemize}[nosep]
\item $\Phi$ (scatter-add): global accumulator array of $N$ scores---$O(N)$ working set, cache-hostile.
\item $\Psi$ (window-switch): windowed accumulator of size $W$---$O(W)$ working set, cache-friendly. Used throughout the paper.
\item $\Xi$ (SIMD multiply): same global array as $\Phi$ with SIMD vectorization --- tests whether arithmetic throughput matters.
\end{itemize}
Linux perf \cite{de2010new} on MS~MARCO+SPLADE (Table~\ref{tab:perf_engines}, top rows) confirms: the locality-aware $\Psi$ achieves 2.4$\times$ latency reduction by virtually eliminating cache and TLB misses (2.7\% vs.\ 30\%), while the SIMD-focused $\Xi$ provides no measurable benefit (Figure~\ref{fig:profiling})---confirming the workload is \emph{memory-bound, not compute-bound}.

\paragraph{Connection to complementarity.}
Static pruning reduces \emph{memory traffic} (shorter posting lists, smaller working set); dynamic pruning reduces \emph{wasted traversal} (skipping non-competitive blocks/clusters).
Table~\ref{tab:perf_engines} extends the profiling to BMP and SEISMIC: per-query latency tracks absolute cache misses on both engines, and configurations with smaller indexes consistently exhibit fewer misses per query---confirming that index-side pruning reduces memory traffic across all tested architectures.
SEISMIC's ${\sim}$17$\times$ lower cache misses per query (6.5K vs.\ BMP's 114K) stem primarily from candidate \emph{selection}, not cheaper per-document access: at qc=5 it scores only the 5 highest-weight query terms' clusters versus all ${\sim}$44 SPLADE terms traversed by the exhaustive pipeline and BMP---an 8.8$\times$ reduction in candidate breadth. Normalized per query-term list, SEISMIC (1.3K misses/list) and the exhaustive $\Psi$ (1.5K) differ by only ${\sim}$1.2$\times$, so the cluster-contiguous layout is second-order; within SEISMIC, varying only \texttt{query\_cut} on a fixed access path, misses scale near-linearly with active terms ($r$=0.992)---confirming selection, not dot-product implementation, as the dominant lever.

\begin{table}[h]
\centering
\caption{Micro-architectural profile across engines (MS~MARCO+SPLADE). Cache-miss and dTLB-miss are rates; CacheMiss/Q is the absolute count per query driving latency.}
\label{tab:perf_engines}
\small
\setlength{\tabcolsep}{2.5pt}
\begin{tabular}{@{}l l r r r r@{}}
\toprule
\textbf{Engine} & \textbf{Config} & \textbf{IPC} & \textbf{Cache\%} & \textbf{dTLB\%} & \textbf{CacheMiss/Q} \\
\midrule
C++ $\Phi$ & scatter-add & 1.58 & 30.0 & 97.0 & 740{,}000 \\
C++ $\Psi$ & window-switch & 2.50 & 2.7 & 2.6 & 67{,}000 \\
\midrule
BMP & $\beta$=1.0 & 3.37 & 1.3 & 28.4 & 114{,}000 \\
BMP & $\beta$=0.5 & 3.37 & 1.8 & 25.0 & 89{,}000 \\
\midrule
SEISMIC & qc=20 & 3.01 & 10.9 & 40.9 & 12{,}500 \\
SEISMIC & qc=5 & 3.29 & 7.7 & 37.7 & 6{,}500 \\
\bottomrule
\end{tabular}
\end{table}


\section{How Far to Push: Operating Point Selection}
\label{sec:metrics}

\textbf{RQ1} established \emph{what} to prune (documents: portable; queries: regime-dependent) and \textbf{RQ2} showed \emph{how} to combine pruning mechanisms (static and dynamic are complementary).
The remaining deployment question \textbf{RQ3} is \emph{how aggressively} to prune.
We show that NDCG@10 provides a portable stopping criterion across engines.

\subsection{NDCG@10 Saturates Before Recall@10---Portably}

Across all three engines and both datasets, NDCG@10 plateaus while Recall@10 continues to change.
On all of them, NDCG@10 remains within 0.008 of baseline while (oracle) Recall@10 drops by up to 14 points (Figure~\ref{fig:ndcg_saturation})---i.e.\ NDCG saturates while Recall is still in the high-0.8 to mid-0.9 range, long before the aggressive-pruning regime.
For example, BMP retains NDCG=0.449 at Recall=0.932 ($\beta$=0.5). Crucially, the saturation replicates under the \emph{deep, graded} judgments of TREC~DL~2019/2020 (${\sim}$210 judged docs/query): index-side pruning retaining ${\sim}$90\% mass stays within 0.006 NDCG@10 of the exact baseline on all three engines (Appendix~\ref{app:dl}), so it is not an artifact of MS~MARCO's sparse relevance labels.
This convergence across architecturally distinct engines suggests the saturation is a \emph{robust property} of learned sparse retrieval under qrels-based evaluation, not an engine-specific artifact.

Success@10 saturates later than NDCG but earlier than Recall, remaining above 0.96 on BMP across the full $\beta$ sweep---confirming that moderate pruning preserves candidate-generation viability.
On MS~MARCO dev, MRR@10 tracks NDCG@10 in lockstep across every engine and operating point (MRR@10 columns in Tables~\ref{tab:controlled_ops},~\ref{tab:bmp_factorial},~\ref{tab:seismic_factorial} and Appendix~\ref{app:detailed}): index-side pruning at the saturation point preserves it while only aggressive pruning degrades it, so the saturation is not an NDCG-specific artifact.

\subsection{Root Cause and Stopping Point}

\begin{figure*}[h]
    \centering
    \includegraphics[width=0.85\textwidth]{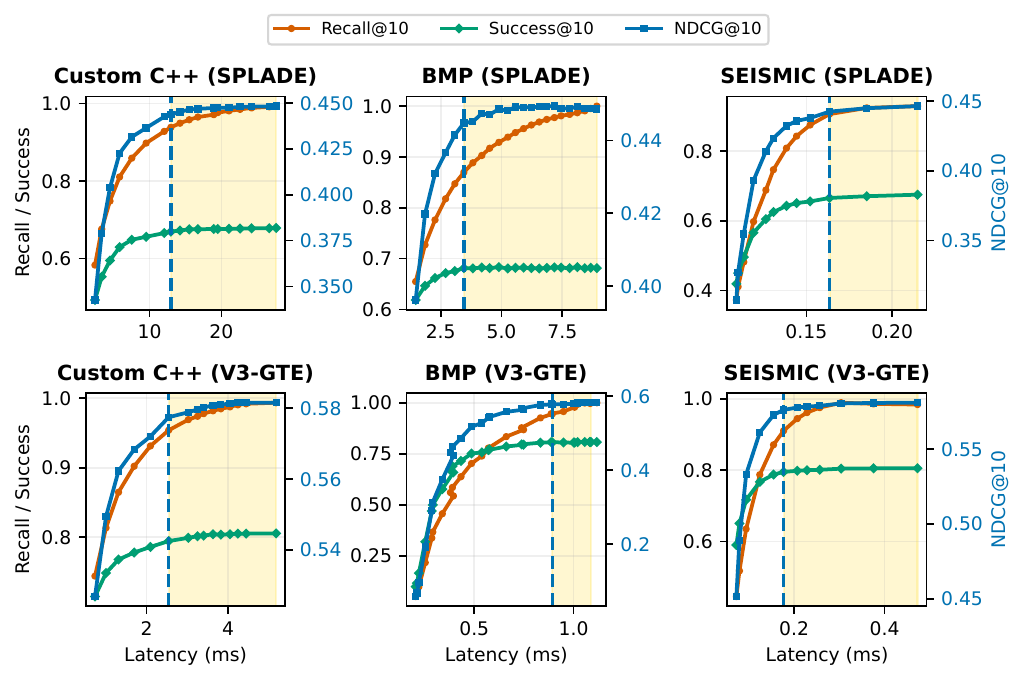}
    \caption{Latency vs.\ quality across three engines (columns) and two dataset--encoder regimes (rows: MS~MARCO+SPLADE, top; NQ+V3-GTE, bottom). The dashed line marks the NDCG@10 knee (the most aggressive point within 0.005 of the unpruned baseline); the gold-shaded region to its right is where NDCG@10 has saturated while Recall@10 keeps rising, confirming the metric gap is engine-independent.}
    \label{fig:ndcg_saturation}
\Description{Six panels: three engines (columns) by two dataset--encoder regimes (rows). A dashed line marks the NDCG knee in each panel; the gold-shaded region to its right highlights where NDCG has saturated while Recall keeps rising.}
\end{figure*}

The saturation arises because pruning tends to swap top-ranked documents for near-duplicate passages of \emph{equal} relevance grade (or for unjudged near-duplicates): the graded-gain vector over ranks 1--10 is left essentially invariant even though the exact top-$k$ set (and thus oracle Recall) changes, so NDCG@10 is blind to the reshuffle (verified on a deep-judgment DL~2019 query where all churned top-10 documents are judged equally relevant; Appendix~\ref{app:dl}).
We define the NDCG knee as the most aggressive operating point whose NDCG@10 stays within 0.005 of the unpruned baseline (a ${\sim}$1\% relative tolerance), marked by the dashed line in Figure~\ref{fig:ndcg_saturation}. Across all three engines and both encoders, on dev and the deep-judgment DL~2019/2020 sets, this knee falls where oracle Recall@10 lies in the ${\sim}$85--95\% range (engine-dependent).
Depending on the deployment objective, practitioners should select operating points as follows:
\begin{enumerate}[nosep]
    \item \textbf{System profiling}: Use Recall@10---it remains sensitive even in the high-accuracy regime (e.g., 0.97 vs.\ 0.99), diagnosing approximation fidelity.
    \item \textbf{User-facing ranking}: Target the NDCG@10 knee (${\sim}$85--95\% Recall, engine-dependent). On BMP, $\beta$=0.5 (NDCG=0.449, Recall=0.93) is near-optimal; on SEISMIC, qc=5 (NDCG=0.443, Recall=0.98) operates in the saturated region.
    \item \textbf{Candidate generation}: Use Success@10 to ensure coverage of judged-relevant items; at doc AM~0.5 on the controlled pipeline, Success@10 $\ge$ 0.95 suffices for downstream re-ranking.
\end{enumerate}
The NDCG knee is \emph{portable across all three engines}, completing the decision framework: \emph{what} to prune (RQ1: documents), \emph{how} to combine (RQ2: static + dynamic), and \emph{how far} to push (RQ3: to the NDCG knee at ${\sim}$85--95\% Recall).

\section{Discussion and Practitioner Guidance}
\label{sec:discussion}

\paragraph{Decision framework: What, How, How Far.}
Our cross-engine results distill into a deployment workflow:
\begin{enumerate}[nosep]
    \item \textbf{Start with document pruning} (Doc AM\,0.90; tune the system to $\Delta$NDCG $\le$ 0.005).
    \item \textbf{Use built-in query reduction} (BMP $\beta$, SEISMIC qc); external query pruning is secondary.
    \item \textbf{Add posting-list pruning when RAM is the binding constraint}; tune the threshold under a validation constraint, as Recall degrades faster than document pruning on some engine/encoder combinations.
    \item \textbf{Combine under tight latency budgets} (BMP 2.52$\times$, SEISMIC 1.39$\times$).
\end{enumerate}

\paragraph{Limitations.}
The controlled C++ pipeline uses two-stage re-ranking while BMP and SEISMIC are single-stage; Table~\ref{tab:portability_matrix} therefore reports a C++$^\dagger$ no-rerank variant, which preserves the qualitative portability conclusion but lowers absolute Recall.
Our portability claims are established across three engine paradigms (exhaustive scoring, block-max dynamic pruning, and clustered inverted indexes), two learned-sparse encoders spanning opposite query-density regimes (SPLADE ${\sim}$44 vs.\ V3-GTE ${\sim}$7 query terms), two web-search benchmarks (MS~MARCO, NQ) plus the deep-judgment TREC~DL~2019/2020 sets, and $k$=10 under single-threaded execution. We do not claim generality to all learned-sparse models, non-web domains (broader BEIR coverage is bounded by CPU-only encoding cost and left to future work), batched/multi-threaded settings, deeper $k$, or additional engines (DSP, SINDI). LLM-as-a-judge evaluation is an alternative-validity path we do not pursue, as the deep DL judgments already directly test the shallow-label concern.
The saturation point is a qrels-based operating criterion (validated under both shallow dev and deep DL judgments), not a direct user-satisfaction threshold.

\section{Conclusion}
\label{sec:conclusion}

Across 1{,}140 configurations on three engines, two datasets, and two encoders, we find that index-side pruning is portable (1.2--6.6$\times$ speedup, 18--82\% index reduction), query pruning is subsumed by engine-internal mechanisms (BMP's $\beta$, SEISMIC's \texttt{query\_cut}), and static pruning complements dynamic pruning (combined 2.52$\times$ on BMP).
The NDCG knee at ${\sim}$85--95\% Recall provides a portable empirical stopping criterion that replicates under deep TREC~DL judgments.
Micro-architectural profiling confirms all tested engines are memory-traffic-limited, explaining why index-size reduction transfers across architectures.
A no-rerank ablation further shows that the cross-engine recall gap reflects traversal semantics rather than re-ranker protection.
Future work should extend to $k$=100+, additional engines (DSP, SINDI), impact quantization interactions, and multi-threaded execution.

\appendix

\section{Accumulator Variants}
\label{app:accumulators}

We implement three accumulator strategies in the controlled C++ pipeline to isolate the memory-bound bottleneck. The key difference is working-set size: $\Phi$ maintains a global accumulator of $N$ scores (random access, cache-hostile); $\Psi$ partitions the document space into windows of size $W$ (localized, cache-friendly); $\Xi$ adds SIMD vectorization to $\Phi$'s multiply step, yielding $<$1\% improvement---confirming the workload is memory-bound, not compute-bound. All controlled-pipeline experiments use $\Psi$ unless noted.

\vspace{2pt}
\noindent\fbox{\parbox{0.97\linewidth}{\small
Both score query $q$ against inverted index $I$ then re-rank the top-$k'$ with full dot products; they differ only in the accumulator.\\[2pt]
\textbf{$\Phi$ Scatter-Add (baseline).} One global array $S[0\ldots N{-}1]$: \textbf{for} each term $(t,q_t)$, \textbf{for} each posting $(d,w_d)\in I[t]$: $S[d]\mathrel{+}= q_t w_d$. \textcolor{gray}{Working set $O(N){\approx}$33\,MB.}\\[2pt]
\textbf{$\Psi$ Window-Switch (used throughout).} \textbf{for} each window $w$ of $W$ docs: reset local $S[0\ldots W{-}1]$, accumulate as above over $I[t][w]$, then push the window's positives to a global heap. \textcolor{gray}{Working set $O(W){\approx}$400\,KB, cache-resident.}}}

\section{Detailed Experimental Results}
\label{app:detailed}
\renewcommand{\arraystretch}{0.9}

Tables~\ref{tab:bmp_doc_detail} and~\ref{tab:seismic_doc_detail} report the full per-configuration BMP and SEISMIC sweeps (index size, latency, Recall@10, NDCG@10, MRR@10) underlying the main portability findings; controlled-pipeline operating points appear in Table~\ref{tab:controlled_ops}.

\begin{table}[htbp]
\centering
\caption{BMP document/posting/query pruning across all four dataset--encoder combinations ($\alpha$=1.0). Speedup relative to each config's unpruned baseline.}
\label{tab:bmp_doc_detail}
\small
\setlength{\tabcolsep}{2.5pt}
\resizebox{\linewidth}{!}{%
\begin{tabular}{llrrrrrr}
\toprule
Config & Pruning & Idx (GB) & Lat ($\mu$s) & R@10 & NDCG & MRR & Spdup \\
\midrule
\multirow{6}{*}{\rotatebox{0}{\scriptsize MS+SPL}}
 & None & 15.7 & 8{,}915 & 1.000 & 0.449 & 0.382 & 1.00$\times$ \\
 & AM 0.50 & 4.3 & 2{,}869 & 0.685 & 0.398 & 0.335 & 3.11$\times$ \\
 & AM 0.70 & 6.4 & 4{,}166 & 0.832 & 0.431 & 0.366 & 2.14$\times$ \\
 & AM 0.90 & 10.1 & 6{,}235 & 0.948 & 0.446 & 0.380 & 1.43$\times$ \\
 & AM 0.95 & 11.7 & 7{,}111 & 0.972 & 0.448 & 0.381 & 1.25$\times$ \\
 & MR 0.10 & 10.4 & 6{,}281 & 0.951 & 0.446 & 0.380 & 1.42$\times$ \\
\midrule
\multirow{6}{*}{\rotatebox{0}{\scriptsize NQ+SPL}}
 & None & 5.5 & 4{,}272 & 1.000 & 0.539 & 0.489 & 1.00$\times$ \\
 & AM 0.50 & 1.5 & 1{,}412 & 0.624 & 0.454 & 0.407 & 3.02$\times$ \\
 & AM 0.70 & 2.3 & 2{,}277 & 0.800 & 0.518 & 0.470 & 1.88$\times$ \\
 & AM 0.90 & 3.6 & 3{,}222 & 0.936 & 0.537 & 0.488 & 1.33$\times$ \\
 & AM 0.95 & 4.1 & 3{,}573 & 0.968 & 0.538 & 0.488 & 1.20$\times$ \\
 & MR 0.10 & 3.7 & 3{,}336 & 0.951 & 0.535 & 0.487 & 1.28$\times$ \\
\midrule
\multirow{4}{*}{\rotatebox{0}{\scriptsize MS+GTE}}
 & None & 22.4 & 2{,}453 & 1.000 & 0.429 & 0.363 & 1.00$\times$ \\
 & $\beta$=0.5 & 22.4 & 1{,}061 & 0.706 & 0.365 & 0.302 & 2.31$\times$ \\
 & Doc AM~0.90 & 14.4 & 1{,}994 & 0.969 & 0.428 & 0.362 & 1.23$\times$ \\
 & Post MR~0.10 & 13.8 & 1{,}934 & 0.969 & 0.427 & 0.361 & 1.27$\times$ \\
\midrule
\multirow{4}{*}{\rotatebox{0}{\scriptsize NQ+GTE}}
 & None & 7.1 & 1{,}087 & 1.000 & 0.583 & 0.536 & 1.00$\times$ \\
 & $\beta$=0.5 & 7.1 & 571 & 0.778 & 0.543 & 0.498 & 1.90$\times$ \\
 & Doc AM~0.90 & 4.7 & 917 & 0.966 & 0.579 & 0.532 & 1.19$\times$ \\
 & Post MR~0.10 & 4.7 & 890 & 0.966 & 0.578 & 0.531 & 1.22$\times$ \\
\bottomrule
\end{tabular}
}
\end{table}

\begin{figure*}[t]
\centering
\includegraphics[width=0.9\textwidth]{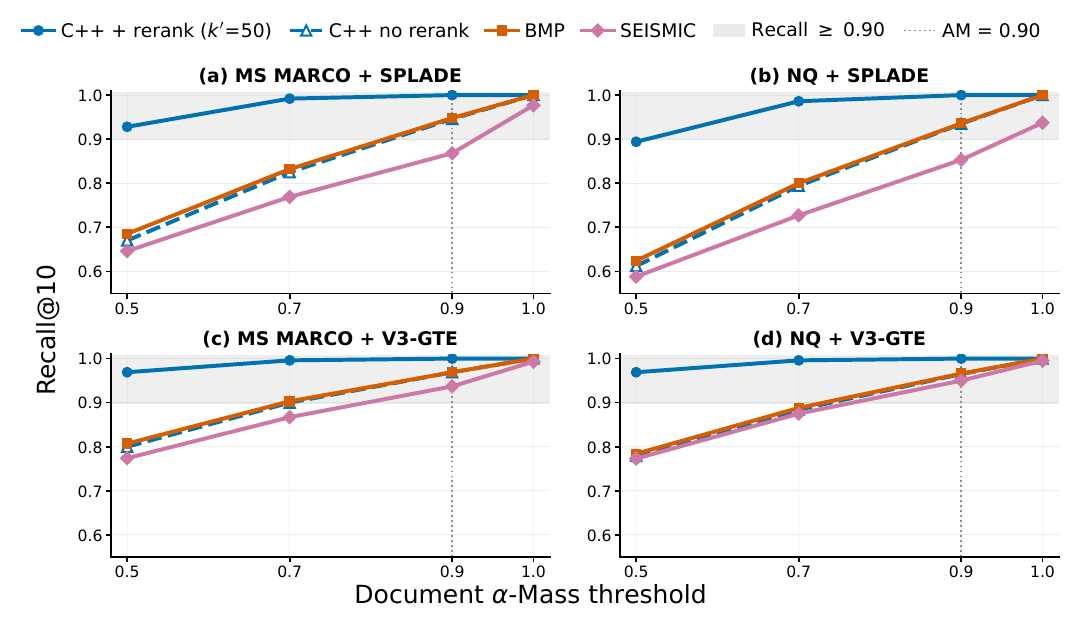}
\caption{Recall@10 under document pruning across the four dataset--encoder combinations.
The two-stage C++ curve shows the effect of full-vector re-ranking, while the no-rerank C++ curve closely tracks BMP at the main operating point (AM=0.90).
SEISMIC exhibits a larger Recall drop under the same static pruning threshold, indicating that pruning thresholds have different effective severity under cluster-based traversal. The gray band marks Recall@10 $\geq 0.90$.}
\label{fig:rerank_ablation}
\Description{Four-panel figure comparing Recall@10 under document alpha-mass pruning for C++ with reranking, C++ without reranking, BMP, and SEISMIC.}
\end{figure*}

\begin{table}[htbp]
\centering
\caption{SEISMIC document pruning across all configurations (qc=5, hf=1.0).}
\label{tab:seismic_doc_detail}
\small
\setlength{\tabcolsep}{2.5pt}
\begin{tabular}{llrrrrr}
\toprule
Config & Pruning & Idx (GB) & Lat ($\mu$s) & R@10 & NDCG & MRR \\
\midrule
\multirow{4}{*}{\rotatebox{0}{\scriptsize MS+SPL}}
 & None & 8.1 & 186 & 0.977 & 0.443 & 0.379 \\
 & AM 0.70 & 4.3 & 140 & 0.769 & 0.424 & 0.361 \\
 & AM 0.90 & 6.0 & 158 & 0.868 & 0.438 & 0.374 \\
 & AM 0.95 & 6.7 & 167 & 0.889 & 0.441 & 0.378 \\
\midrule
\multirow{4}{*}{\rotatebox{0}{\scriptsize NQ+SPL}}
 & None & 7.0 & 234 & 0.937 & 0.533 & 0.485 \\
 & AM 0.70 & 4.5 & 167 & 0.727 & 0.505 & 0.460 \\
 & AM 0.90 & 5.9 & 190 & 0.853 & 0.531 & 0.484 \\
 & AM 0.95 & 6.3 & 204 & 0.880 & 0.534 & 0.486 \\
\midrule
\multirow{4}{*}{\rotatebox{0}{\scriptsize MS+GTE}}
 & None & 10.3 & 255 & 0.993 & 0.422 & 0.358 \\
 & AM 0.70 & 5.0 & 144 & 0.867 & 0.413 & 0.350 \\
 & AM 0.90 & 7.3 & 179 & 0.937 & 0.420 & 0.357 \\
 & AM 0.95 & 8.2 & 207 & 0.951 & 0.420 & 0.356 \\
\midrule
\multirow{4}{*}{\rotatebox{0}{\scriptsize NQ+GTE}}
 & None & 8.3 & 360 & 0.995 & 0.581 & 0.534 \\
 & AM 0.70 & 5.1 & 208 & 0.875 & 0.563 & 0.518 \\
 & AM 0.90 & 6.8 & 239 & 0.950 & 0.574 & 0.528 \\
 & AM 0.95 & 7.4 & 271 & 0.966 & 0.577 & 0.530 \\
\bottomrule
\end{tabular}
\end{table}

\section{Re-Ranking Ablation}
\label{app:rerank_ablation}

The controlled C++ pipeline uses two-stage re-ranking ($k'$=50) while BMP and SEISMIC are single-stage. Figure~\ref{fig:rerank_ablation} shows that at the operating point (Doc AM\,0.90) removing re-ranking costs only 3--6 Recall@10 points---bringing C++ in line with BMP (cf.\ the C++$^\dagger$ column of Table~\ref{tab:portability_matrix})---while at aggressive Doc AM\,0.50 it recovers 17--28 points, confirming its value only beyond the recommended range.

\begin{table}[h!]
\centering
\caption{TREC DL 2019/2020 mean NDCG@10 (DL19+DL20 pooled). Query pruning uses each engine's native mechanism (C++ MR\,0.10; BMP $\beta$=0.5; SEISMIC \texttt{query\_cut}=5)}
\label{tab:dl_validation}
\small
\setlength{\tabcolsep}{3pt}
\begin{tabular}{@{}l ccc c ccc@{}}
\toprule
& \multicolumn{3}{c}{\textbf{SPLADE}} & & \multicolumn{3}{c}{\textbf{V3-GTE}} \\
\cmidrule(lr){2-4} \cmidrule(lr){6-8}
\textbf{Config} & \textbf{C++} & \textbf{BMP} & \textbf{SEIS.} & & \textbf{C++} & \textbf{BMP} & \textbf{SEIS.} \\
\midrule
Baseline        & .726 & .726 & .721 & & .720 & .718 & .721 \\
Doc AM\,0.90    & .717 & .719 & .707 & & .721 & .721 & .718 \\
Post MR\,0.10   & .720 & .723 & .716 & & .720 & .719 & .721 \\
Doc AM\,0.50    & .680 & .691 & .675 & & .688 & .685 & .676 \\
Query prune     & .728 & .728 & .722 & & .723 & .620 & .720 \\
\bottomrule
\end{tabular}
\end{table}

\section{TREC DL 2019/2020 Deep-Judgment Validation}
\label{app:dl}

MS~MARCO dev labels only one or two relevant passages per query, so we re-evaluate on TREC~DL~2019/2020, which pools ${\sim}$210 documents per query and grades each on a 0--3 relevance scale. Table~\ref{tab:dl_validation} confirms both main-text findings under deep judgments: index-side pruning is lossless, and the query-pruning regime split (free on SPLADE, catastrophic on V3-GTE) is if anything sharper. Because these sets judge ${\sim}$210 documents per query rather than one or two, a genuine quality loss would register here; that NDCG@10 stays essentially flat through pruning confirms the saturation in Section~\ref{sec:metrics} is a property of the ranking, not of sparse labels.

\section*{GenAI Usage Disclosure}
Generative AI tools were used only to polish author-written text; they were not used to generate ideas, design experiments, produce or analyze results, write code, or create figures/tables. The authors reviewed all affected text and take full responsibility for this paper.

\bibliographystyle{ACM-Reference-Format}
\balance
\bibliography{references}

\end{document}